\documentstyle[aps,epsf]{revtex}

\begin{document}

\draft

\preprint{HIP-1999-13/TH, DPSU-99-4}

\title{Phenomenological Constraints on 
SUSY $SU(5)$ GUTs with Non-universal Gaugino Masses}

\author{Katri Huitu$^{\text{a}}$, Yoshiharu Kawamura$^{\text{b}}$, 
Tatsuo Kobayashi$^{\text{a}}$ and Kai Puolam{\"{a}}ki$^{\text{a}}$}

\address{$^{\text{a}}$Helsinki Institute of Physics, P.O.Box 9,
FIN-00014 University of Helsinki, Finland }
\address{$^b$ Department of Physics, Shinshu University,
Matsumoto 390-0802, Japan}

\date{September 9, 1999}

\maketitle

\begin{abstract}

We study phenomenological aspects of supersymmetric $SU(5)$ grand
unified theories with non-universal gaugino masses.  For large $\tan
\beta$, we investigate constraints from the requirement of successful
electroweak symmetry breaking, the positivity of stau mass squared and
the $b \to s \gamma$ decay rate.  In the allowed region, the nature of
the lightest supersymmetric particle is determined.  Examples of mass
spectra are given.  We also calculate loop corrections to the bottom
mass due to superpartners.

\end{abstract}

%See <URL:http://www.aip.org/pacs/> for PACS numbers.
% 11.30.Pb       Supersymmetry (see also 12.60.J Supersymmetric
% models)
% 12.10.Dm       Unified theories and models of strong and electroweak
%                interactions

\section{Introduction}

Supersymmetric gauge field theories are among the most promising
models for physics beyond the standard model. The low-energy
supersymmetry (SUSY) solves the so-called hierarchy problem, which
basically follows from the tremendous scale differences in realistic
models including gravity.

After SUSY breaking, SUSY models, e.g. the minimal supersymmetric
standard model (MSSM) have over hundred free parameters in general.
Most of these new parameters in the MSSM are in fact related to the
SUSY breaking, i.e. gaugino masses $M_a$, soft scalar masses $m_i$,
SUSY breaking trilinear couplings $A_{ijk}$ and SUSY breaking bilinear
couplings.  They are expected to be of the order of 1~TeV.

To probe the SUSY breaking mechanisms is very important in order to
produce solid information on physics beyond the standard model. Two
types of SUSY breaking mechanisms, gravity-mediated SUSY breaking and
gauge-mediated SUSY breaking, have been actively studied in recent
years.  The signatures of gravity mediated and gauge mediated SUSY
breaking are quite different. A specific SUSY breaking mechanism
usually reduces the number of {\em a priori} free parameters from
about one hundred to only a few by introducing solid relations among
the SUSY breaking parameters. This makes the phenomenology of the MSSM
more accessible for study.

For phenomenology of SUSY models, various aspects have been studied in
several regions of the parameter space.  Most phenomenological
analyses have been done under the assumption that the soft SUSY
breaking parameters are universal, i.e. $M_a = M_{1/2}$ for $a
=1,2,3$, $m_i=m_0$ for any scalar and $A_{ijk}=A$ at a certain energy
scale, e.g. the Planck scale or the grand unified theory (GUT)
scale. From the phenomenological viewpoint, the universality
assumption is useful to simplify analysis.  Actually, the universal
parameters can be derived from a certain type of underlying theories,
e.g. minimal supergravity.

However, the universality assumption may remove some interesting
degrees of freedom.  Indeed, there exist interesting classes of models
in which non-universal soft SUSY breaking terms can be derived.  For
example, string-inspired supergravity can lead to non-universality for
SUSY breaking parameters at the Planck scale \cite{BIM,st-soft}.
Also, gauge-mediated SUSY breaking models, in general, lead to
non-universality \cite{gmed}.

Recently phenomenological implications of non-universal SUSY breaking
parameters have been investigated.  For example, in
Ref.\cite{nonuni2,KNOY} phenomenological implications have been
studied for non-universal gaugino masses derived from string models.
GUTs without a singlet also lead to non-universal gaugino masses
\cite{EKN,gunion,GLMR}.  In Ref.\cite{gunion} phenomenological aspects
in the small $\tan \beta$ scenario have been discussed, e.g. mass
spectra and some decay modes.  Some phenomenological constraints
reduce the allowed region of the universal SUSY breaking parameters a
lot.  For example, in the large $\tan \beta$ scenario, it is hard to
fulfill the constraints due to the requirement of successful
electroweak breaking, SUSY corrections to the bottom mass
\cite{hall,COPW} and the $b \to s \gamma$ decay rate
\cite{bertolini,bs}.  These constraints can be relaxed in
non-universal cases.

In this paper, we study phenomenological aspects of SUSY $SU(5)$ GUTs
where the gaugino masses come from a condensation of $F$-component
with a representation {\bf 24}, {\bf 75} or {\bf 200}.  Each of them
leads to a proper pattern of non-universal gaugino masses.  We mostly
concentrate on the large $\tan \beta$ scenario.  We take into account
the full one-loop effective potential of the MSSM, in order to
calculate the physical spectrum of the MSSM, given the initial
conditions at the GUT scale. In particular, we investigate constraints
from the requirement of successful electroweak symmetry breaking, the
positivity of stau mass squared and the $b \to s \gamma$ decay rate.
We take SUSY corrections to the bottom quark mass carefully into
account.  We then find the allowed parameter space for each model and
describe the particle spectrum.

This paper is organized as follows.  In section 2 SUSY $SU(5)$ GUTs
with non-universal gaugino masses are reviewed.  In section 3 we study
their phenomenological aspects, i.e. successful radiative breaking of
the electroweak symmetry, the LSP mass, the stau mass, SUSY
corrections to the bottom mass and the $b \to s \gamma$ decay.  We
also give comments on small $\tan \beta$ cases.  Section 4 is devoted
to conclusions.

\section{SUSY $SU(5)$ GUTs with non-universal gaugino masses}

We discuss the non-universality of soft SUSY breaking gaugino masses
in SUSY $SU(5)$ GUT and the constraints on parameters at the GUT scale
$M_X$ in our analysis.  The gauge kinetic function is given by
\begin{eqnarray}
{\cal L}_{g.k.} &=& \sum_{\alpha, \beta} \int d^2\theta
f_{\alpha\beta}(\Phi^I) W^\alpha W^\beta + H.c.
\nonumber \\
&=& -{1 \over 4} \sum_{\alpha, \beta} Re f_{\alpha\beta}(\phi^I)
F^\alpha_{\mu\nu} F^{\beta \mu\nu} + \sum_{\alpha, \beta, \alpha',
\beta'} \sum_I F^I_{\alpha'\beta'} {\partial f_{\alpha\beta}(\phi^I)
\over \partial \phi^I_{\alpha'\beta'}}
\lambda^\alpha \lambda^\beta   +   H.c. + \cdots
\label{gaugekinetic}
\end{eqnarray}
where $\alpha, \beta$ are indices related to gauge generators,
$\Phi^I$'s are chiral superfields and $\lambda^\alpha$ is the $SU(5)$
gaugino field.  The scalar and $F$-compontents of $\Phi^I$ are denoted
by $\phi^I$ and $F^I$, respectively.  The $\Phi^I$'s are classified
into two categories.  One is a set of $SU(5)$ singlet supermultiplets
$\Phi^S$ and the other one is a set of non-singlet ones $\Phi^N$.  The
gauge kinetic function $f_{\alpha\beta}(\Phi^I)$ is, in general, given
by
\begin{eqnarray}
f_{\alpha\beta}(\Phi^I) = f_0(\Phi^S) \delta_{\alpha\beta} 
+ \sum_N \xi_N(\Phi^S) {\Phi^N_{\alpha\beta} \over M}
+ O(({\Phi^N_{\alpha\beta} \over M})^2)
\label{f}
\end{eqnarray}
where $f_0$ and $\xi_N$ are functions of gauge singlets $\Phi^S$ and
$M$ is the reduced Planck mass defined by $M \equiv
M_{Pl}/\sqrt{8\pi}$.  Since the gauge multiplets are in adjoint
representation, one finds the possible representations of $\Phi^N$
with non-vanishing $\xi_N$ by decomposing the symmetric product ${\bf
24} \times {\bf 24}$ as
\begin{equation}
( {\bf 24} \times {\bf 24} )_s = {\bf 1} + {\bf 24}+{\bf 75}+{\bf 200}.
\end{equation}
Thus, the representations of $\Phi^N$ allowed as a linear term of
$\Phi^N$ in $f_{\alpha\beta}(\Phi^I)$ are {\bf 24}, {\bf 75} and {\bf
200}.

Here we make two basic assumptions.  The first one is that SUSY is
broken by non-zero VEVs of $F$-components $F^{I'}$, i.e., $\langle
F^{I'} \rangle = O(m_{3/2}M)$ where $m_{3/2}$ is the gravitino mass.
The second one is that the $SU(5)$ gauge symmetry is broken down to
the standard model gauge symmetry $G_{SM} = SU(3) \times SU(2) \times
U(1)$ by non-zero VEVs of non-singlet scalar fields $\phi^N$ at the
GUT scale $M_X$.

After the breakdown of $SU(5)$, the gauge couplings $g_a$'s of
$G_{SM}$, are, in general, non-universal at the scale $M_X$
\cite{g-nonuni} as we see from the formula $g_a^{-2}(M_X) \delta_{ab}
= \langle Re f_{ab} \rangle$.  The index $a (= 3,2,1)$ represents
$(SU(3), SU(2), U(1))$ generators as a whole.  The gaugino field
acquires soft SUSY breaking mass after SUSY breaking.  The mass
formula is given by
\begin{eqnarray}
M_a(M_X) \delta_{ab} = \sum_I {\langle F^I_{a'b'} \rangle \over 2}
{\langle \partial f_{ab}/\partial \phi^I_{a'b'} \rangle \over 
\langle Re f_{ab} \rangle}  .
\label{Ma}
\end{eqnarray}
Thus the $M_a$'s are also, in general, non-universal at the scale
$M_X$
\cite{EKN}.

Next we will consider the constraints on the physical parameters used
at the scale $M_X$ for the analysis in this paper.\\ 1) Gauge
couplings\\ We take a gauge coupling unification scenario within the
framework of the MSSM, that is,
\begin{eqnarray}
\alpha_1(M_X) = \alpha_2(M_X) = \alpha_3(M_X) \equiv \alpha_X \sim 1/25
\label{gaugecoupling}
\end{eqnarray}
where $\alpha_a \equiv g_a^2/4\pi$ and $M_X = 2.0 \times 10^{16}$GeV.
The relation (\ref{gaugecoupling}) leads to $\langle Re f_0 \rangle
\sim 2$.  We neglect the contribution of non-universality to the gauge
couplings.  Such corrections of order $O(\langle \phi^N \rangle/M) =
O(M_X/M)=O(1/100)$ have little effects on phenomenological aspects
which we will discuss in the next section, although such corrections
would be important for precision study on the gauge coupling
unification.\\ 2) Gaugino masses\\ We assume that dominant component
of gaugino masses comes from one of non-singlet $F$-components.  The
VEV of the $F$-component of a singlet field whose scalar component
$\phi^{S'}$ has a VEV of $O(M)$ in $f_{\alpha\beta}$ is supposed to be
small enough $\langle F^{S'} \rangle \ll O(m_{3/2} M)$ such as dilaton
multiplet in moduli-dominant SUSY breaking in string models.  In this
case, ratios of gaugino masses at $M_X$ are determined by group
theoretical factors and shown in table~I.  The patterns of gaugino
masses which stem from $F$-term condensation of {\bf 24}, {\bf 75} and
{\bf 200} are different from each other.  The table also shows
corresponding ratios at the weak scale $M_Z$ based on MSSM.  In the
table, gaugino masses are shown in the normalization $M_3(M_X)=1$. 
Note that the signs of $M_a$ are also fixed by group theory 
up to an overall phase as shown in Table 1.
There is no direct experimental constraint on these signs. 
For example, these signs affect radiative corrections of 
$A$-terms and thus off-diagonal elements of sfermion matrices, that is, 
radiative corrections of $M_a$ to $A_t$ are constructive 
in the universal case, while 
in the model {\bf 24} radiative corrections between $M_3$ and the others 
interfere with each other leading to $30 \%$ reduction.
In the other cases, the radiative corrections are larger by $20 \sim 30 \%$ 
than the universal case.\\
\\
3) Scalar masses\\ For simplicity, we assume universal soft SUSY
breaking scalar masses $m_0^{\text{GUT}}$ at $M_X$ in our analysis in
order to clarify phenomenological implications of non-universal
gaugino masses.  The magnitude of $m_0^{\text{GUT}}$ is supposed not
to be too large compared with that of $M_a$'s in order not to
overclose the universe with a huge amount of relic abundance of the
lightest neutralino \cite{KNOY}.

The non-universal gaugino masses $M_a$ and scalar masses $m_k$ may
have sizable SUSY threshold corrections for running of gauge couplings
\cite{guni}.  These threshold effects and non-universal contributions
of $O(\langle \phi^N \rangle/M)$ in $g_a^{-2}$ will be discussed
elsewhere.

\section{Phenomenological constraints and mass spectra}

In this section, we study several phenomenological aspects of SUSY
$SU(5)$ GUTs with non-universal gaugino masses.  The patterns of the
gaugino masses in the models are different from each other as shown in
table~I.  That leads to different phenomenology in these models.  For
example, in the model {\bf 24} we have a large gap between $M_1(M_Z)$
and $M_3(M_Z)$, i.e. $M_1(M_Z)/M_3(M_Z) \approx 0.1$. In the model
{\bf 75} gaugino masses are almost degenerate at the weak scale.  In
the model {\bf 200}, $M_2(M_Z)$ is smallest.  Some phenomenological
aspects have been previously studied in the case with low $\tan
\beta$ \cite{gunion}. 
We will study the case of a large value of $\tan \beta$, 
e.g. $\tan \beta\sim 40$.

We take the trilinear scalar couplings, the so-called $A$-terms, to
vanish at the GUT-scale. Similarly, the case with non-vanishing
$A$-terms can be studied, but the conclusions remain qualitatively
unchanged.  Also we ignore the supersymmetric $CP$-violating phase of
the bilinear scalar coupling of the two Higgs fields, the so-called
$B$-term.  Assuming vanishing $A$-terms and a real $B$-term, we have
no SUSY-CP problem.  Ignoring the complex phases has no significant
effect on the results of this work, although they would naturally be
very relevant to the problem of CP-violation.  We could fix magnitudes
of the supersymmetric Higgs mixing mass $\mu$ and $B$ by assuming some
generation mechanism for the $\mu$-term.  However, we do not take such
a procedure here.  We will instead fix these magnitudes by use of the
minimization conditions of the Higgs potential as shall be shown.

Given the quantum numbers of $F^N$ irreducible representation, one can
characterize the models as a function of four parameters: $\tan
\beta$, the gluino mass $M_3^{\text{GUT}}$ at the GUT scale, the universal
mass of the scalar fields ${{m}}_0^{\text{GUT}}$ at the GUT scale and
the sign of the $\mu$-term ${\text{sign}}(\mu)$.

We will check the compatibility of the model with the experimental
branching ratio $b \to s \gamma$.  Since this branching ratio
increases with $\tan \beta$, we will study the four models at the
region of large $\tan \beta$, taking $\tan \beta=40$ as a
representative value and scanning over the gaugino mass and the scalar
mass squared term. We require that the gauge coupling constants unify
at the scale $2.0 \times 10^{16}$GeV.

Successful electroweak symmetry breaking is an important constraint.
The one-loop effective potential written in terms of the VEVs,
$v_u=\langle H_u^0 \rangle$ and $v_d=\langle H_d^0 \rangle$, is
\begin{equation}
V(Q)=V_0(Q)+\Delta V(Q) ,
\end{equation}
where
\begin{eqnarray}
V_0(Q) & = & \left( m_{H_d}^2+\mu^2 \right) v_d^2 + \left(
m_{H_u}^2+\mu^2 \right) v_u^2 -2 B v_u v_d+\frac 18 \left( g^2+g'^2
\right) \left(v_u^2-v_d^2 \right)^2 , \nonumber \\ \Delta V(Q) & = &
\frac 1{64 \pi^2} \sum_{k=\text{all the MSSM fields}} \left( -1
\right)^{2 S_k} n_k M_k^4 \left[ -\frac 32+\ln \frac{M_k^2}{Q^2}
\right] ,
\label{eq:potform}
\end{eqnarray}
where $S_k$ and $n_k$ are respectively the spin and the number of
degrees of freedom.  Here $m_{H_u}$ and $m_{H_d}$ denote soft SUSY
breaking Higgs masses.

We use the minimization conditions of the full one-loop effective
potential,
\begin{equation}
\frac{\partial V}{\partial v_u} = \frac{\partial V}{\partial v_d}
=0,
\label{eq:mineq0}
\end{equation}
so that we can write $\mu^2=\mu_0^2+\delta \mu^2$ and $B=B_0+\delta B$
in terms of other parameters, that is, the soft scalar masses, the
gaugino masses and $\tan \beta$.  Here $\mu_0^2$ and $B_0$ denote the
values determined only by use of the tree-level potential, and $\delta
\mu^2$ and $\delta B$ denote the corrections due to the full one-loop
potential, which are obtained
\begin{eqnarray}
\delta\mu^2 & = & \frac 12 \frac{v_d \partial \Delta V /\partial
v_d-v_u \partial \Delta V/\partial v_u}{v_u^2-v_d^2} , \nonumber \\
\delta B & = & \frac 12 \frac{v_u \partial \Delta V /\partial v_d-v_d
\partial \Delta V/\partial v_u}{v_u^2-v_d^2} .
\label{eq:potcorr}
\end{eqnarray}
Numerically, the most significant one-loop contribution to $\delta B$
and $\delta \mu^2$ comes from the (s)top and (s)bottom
loops~\cite{kaiandotherfriends}.

Successful electroweak symmetry breaking requires $\mu^2 > 0$.
Furthermore, we require the mass squared eigenvalues for all scalar
fields to be non-negative. In particular, in the large $\tan \beta$
scenario the stau mass squared becomes easily negative due to large
negative radiative corrections from the Yukawa coupling against
positive radiative corrections from the gaugino masses.

These constraints are shown in Fig.1.  In the model {\bf 1} with the
universal gaugino mass, requirement of proper electroweak symmetry
breaking excludes the region with very small ($\sim 100$ GeV) scalar
mass and gaugino masses.  In the model ${\bf 24}$ the region where
radiative symmetry breaking fails is considerably larger than in the
model {\bf 1} because of negative $m_{\tilde \tau}^2$ for small
$m_0^{\text{GUT}}$.  In the model ${\bf 24}$ $M_2(M_Z)$ and $M_1(M_Z)$
are quite small compared with $M_3(M_Z)$.  Such small values of
$M_2(M_Z)$ and $M_1(M_Z)$ are not enough to push up $m_{\tilde
\tau}^2$ against large negative radiative correction due to the Yukawa
coupling.  It is interesting to note that in the model {\bf{75}} large
gaugino masses drive the Higgs boson mass-squared $m_{H_{u,d}}^2$ to
very large positive values at the SUSY scale. This aspect combined
with the contribution from the effective potential correction makes
$\mu^2$ small and negative at large gaugino masses. As a result, in
the model {\bf{75}} there are no consistent solutions having large
gluino mass $M_3^{\text{GUT}} \agt 800$ GeV.  Furthermore, around the
border to the region with $\mu^2 <0$, i.e. $M_3^{\text{GUT}} \sim 800$
GeV, the magnitude of $|\mu|$ is very small, and the lightest
neutralino and the lighter chargino are almost higgsinos.  Thus, the
region around the border $M_3^{\text{GUT}} \sim 800$ GeV is excluded
by the experimental lower bound of the chargino mass, $m_{\chi^\pm}
\agt 90$ GeV.  The region with $M_3^{\text{GUT}} < 700$ GeV leads to
large $|\mu|$ enough to predict $m_{\chi^\pm} > O(100)$ GeV.  In the
model {\bf 200} radiative symmetry breaking works for all the scanned
values.

From the experimental point of view, a crucial issue is the nature of
the LSP, since it is a decisive factor in determing signals of the
models in detectors.  One candidate for the LSP is the lightest
neutralino $\chi^0$.  In the large $\tan \beta$ scenario, the lightest
stau is another possibility \footnote{In Ref.\cite{stau} cosmological
implications of the stau LSP have been discussed.}.  Figs. 1 show what
is the LSP for the four models.  They also show the excluded region by
the current experimental limit $m_{\tilde \tau_1} \ge 72$
GeV~\cite{LEP2}. The limit on the stau mass excludes the models
{\bf{1}} and {\bf{24}} having small scalar masses. The models
{\bf{75}} and {\bf{200}}, on the other hand, always have relatively
heavy stau, independent of the SUSY parameters, and in the latter two
models the neutralino is always the LSP.  For the models {\bf{1}} and
{\bf{24}}, the content of the LSP is similar and narrow regions lead
to the stau LSP.

In our models the present experimental lower bound of the Higgs mass
does not provide a strong constraint, because in the large $\tan
\beta$ scenario the Higgs mass is heavy.

We also consider the constraint due to the $b\rightarrow s \gamma$
decay.  The prediction of the $b\rightarrow s \gamma$ decay branching
ratio~\cite{bertolini} should be within the current experimental
bounds \cite{cleo}
\begin{equation}
1.0<10^4\times BR_{\text{EXP}}(b \rightarrow s \gamma ) < 4.2.
\end{equation}
Combined with the theoretical uncertainty in the SM prediction
($10^4\times BR_{\rm SM}(b \rightarrow s \gamma )=3.5 \pm 0.3$ ) the
branching ratio must be between 0.3 and 1.4 times the SM prediction.

As expected, the constraint is very strong for negative mu-term
${\text{sign}}(\mu)=-1$~\footnote{We follow the conventional
definition of the sign of $\mu$ \cite{HK}.}, because the
supersymmetric contributions interfere constructively to the
amplitude, causing the branching ratio to exceed the experimental
bound. Figs.~1 show excluded regions due to $BR(b \to s \gamma)$ for
the four models with ${\text{sign}}(\mu)=-1$.  We have taken into
account squark mixing effects. These excluded regions are similar for
the four models.  The regions with small gluino masses
$M_3^{\text{GUT}} \alt 700$ GeV are ruled out due to too large
$b\rightarrow s \gamma$ branching ratio in the four models unless
$m_0^{\text{GUT}} \agt 1$ TeV.  In addition, the model {\bf 75} has an
excluded region with $M_3^{\text{GUT}} \agt 800$ GeV due to the
unsuccessful electroweak symmetry breaking.  Thus, in the model {\bf
75} with ${\text{sign}}(\mu)=-1$ only a narrow region for
$M_3^{\text{GUT}}$ is allowed for $m_0^{\text{GUT}} \alt 1$ TeV.  In
the case with positive mu-term ${\text{sign}}(\mu)=+1$ the constraints
are much weaker, only some models with small gluino and soft scalar
masses are ruled out due to the charged Higgs contribution.

The superpartner-loop corrections to the bottom-quark Yukawa coupling
become numerically sizable for large $\tan \beta$.  These corrections
are significant for precise prediction of the bottom mass.  Thus, we
also show how these SUSY-corrections to the bottom mass depend on our
models with non-universal gaugino masses.  The threshold effect can be
expressed as \cite{hall}
\begin{equation}
\lambda_b^{\text{MSSM}} (m_{\text{SUSY}}) = \lambda_b^{\text{SM}}
(m_{\text{SUSY}})/\left( (1+\delta_b) \cos \beta \right) ,
\end{equation}
where $\lambda_{b}^{{\text{SM}},{\text{MSSM}}}$ are the bottom quark
Yukawa couplings in the standard model and MSSM, respectively. The
dominant part of the corrections is given by
\begin{equation}
\delta_b = \mu \tan \beta \left[ \frac{2 \alpha_3}{3 \pi} M_3
I(m_{\tilde{b}_1}^2,m_{\tilde{b}_2}^2,M_3^2) + \frac{\lambda_t}{16
\pi^2} A_t \lambda_t I(m_{\tilde{t}_1}^2,m_{\tilde{t}_2}^2,\mu^2)
\right] ,
\label{eq:deltab}
\end{equation}
where $\lambda_t$ is the top Yukawa coupling and 
\begin{equation}
I(x,y,z)=-\frac{xy \ln x/y+yz \ln y/z +zx \ln z/x}{(x-y)(y-z)(z-x)} .
\end{equation}

The sign of $\delta_b$ is the same as the one of $\mu$.  Figs.~2 show
for the four models the regions with $0\%<|\delta_b|<10\%$,
$10\%<|\delta_b|<20\%$ and $20\%<|\delta_b|$.  Most of the allowed
regions in the models {\bf 1} and {\bf 24} lead to
$10\%<|\delta_b|<20\%$ for $\tan \beta =40$, while most of the allowed
region in the model {\bf 75} leads to $0\%<|\delta_b|<10\%$.  In the
model {\bf 200}, small $m_0^{\text{GUT}}$ leads to
$10\%<|\delta_b|<20\%$, while large $m_0^{\text{GUT}}$ leads to
$0\%<|\delta_b|<10\%$.

The large correction to the bottom mass affects the $b-\tau$ Yukawa
coupling unification, which is one of interesting aspects in GUTs.  We
assume the $b-\tau$ Yukawa coupling unification at the GUT scale and
use the experimental value $m_\tau =1.777$ GeV.  Without the SUSY
correction $\delta_b$ we would have $m_b(M_Z)=3.3$ GeV for $\tan \beta
=40$.  The present experimental value of the bottom mass contains
large uncertainties: Ref. \cite{bmass1}, for instance, gives
\begin{equation}
m_b(M_Z)= 2.67 \pm 0.50 \ {\rm GeV}~,
\label{mbex1}
\end{equation}
while the analysis of the $\Upsilon$ system \cite{bmass2} and the
lattice result \cite{bmass3} give $m_b(m_b) = 4.13 \pm 0.06$ GeV and
$4.15 \pm 0.20$ GeV, respectively, \footnote{See also
ref. \cite{bmass4}.} which translate into
\begin{equation}
m_b(M_Z) = 2.8 \pm 0.2 \ {\rm GeV}.
\label{mbex2}
\end{equation}
Thus, the negative SUSY corrections, that is $\mu <0$, with
$10\%<|\delta_b|<20\%$ are favored for $\tan \beta =40$.  Hence most
of the region in the model ${\bf 75}$ leads to too small $|\delta_b|$
to fit the experimental value for $\tan \beta =40$.  The SUSY
correction $\delta_b$ is proportional to $\tan \beta$.  Therefore, in
the case with large $\tan \beta$, e.g. $\tan \beta = 50$ and 55, some
parameter regions in the model {\bf 75} as well as the model {\bf 200}
become more favorable.  Because the prediction $m_b(M_Z)=3.3$ GeV
without the SUSY correction $\delta_b$ is similar for $\tan \beta
=40$, 50 and 55.

Finally we show sparticle spectra in the regions allowed by the
electroweak breaking conditions and the constraint due to $BR(b \to s
\gamma)$ for $\tan \beta =40$ and ${\text{sign}}(\mu)=-1$.  The whole
particle spectrum is fixed by gluino mass $M_3^{\text{GUT}}$, the soft
scalar mass $m_0^{\text{GUT}}$ and $\tan \beta$. The sign of the
$\mu$-term ${\text{sign}} (\mu)$ has numerically insignificant effect
to the mass spectrum.  In the case of negative $\mu$-term the
experimental upper bound to the $b\rightarrow s \gamma$ decay
branching ratio severely restricts the parameter space. As an example,
we show mass spectra of the four models for
$(M_3^{\text{GUT}},m_0^{\text{GUT}})[{\rm GeV}]=(800,400)$ and
$(100,1500)$ in table~II.  These parameters correspond to almost
smallest mass parameters allowed by theoretical and experimental
considerations common in the four models.  Most of the non-SM degrees
of freedom have masses around 1 TeV.  Note that the model {\bf 75}
with $M_3^{\text{GUT}} =800$ GeV and $m_0^{\text{GUT}}=400$ GeV
predicts very small $|\mu|$ and the lightest chargino mass, which is
actually excluded by the experimental lower bound.  On the other hand,
the model {\bf 24} for small $M_3^{\text{GUT}}$ predicts a very small
mass of the lightest neutralino.

In the models {\bf 75} and {\bf 200} the lightest neutralino
$\chi^0_1$ and the lightest chargino $\chi^\pm_1$ are almost
degenerate. This would potentially create a very difficult
experimental setup~\cite{CDG,nonuni2,CDM}. The charginos would be
extremely difficult to detect, at least near the kinematical
production threshold: as the charginos decay practically all of the
reaction energy is deposited into the invisible LSP neutralinos. If
the charginos decay very close to the interaction point, the photon
background would quite effectively hide the signal. The chargino would
be easy to detect only if it is sufficiently stable, having a decay
length of at least millimeters.

In the models {\bf 1} and {\bf 24}, the LSP is almost the bino.  On
the other hand, the wino-like LSP or the higgsino-like LSP can be
realised in the models {\bf 75} and {\bf 200}.  In particular, the
model {\bf 75} has the region around $M_3^{\text{GUT}} \sim 800$ GeV
where the higgsino is very light.  These different patterns of mass
spectra also have cosmological implications, which will be discussed
elsewhere \cite{nextw}.

We have assumed universal soft scalar mass at the GUT scale in order
to concentrate on phenomenological implications of the non-universal
gaugino masses, but we give some comments on non-universal soft scalar
masses.  Certain types of non-universalities can relax the given
constraints.  For example, the non-universality between the stau mass
and the others is important for the constraint $m_{\tilde \tau}^2 >0$
and obviously a large value of the stau mass at the GUT can remove the
excluded region.  For the electroweak symmetry breaking, the
non-universality between the Higgs masses $m_{Hu}$ and $m_{Hd}$ is
interesting and a large difference of $m_{Hd}^2-m_{Hu}^2$ enlarges the
parameter region for the succesful electroweak symmetry breaking.

We give a comment for the small $\tan \beta$ scenario.  For small
$\tan \beta$, the stau (mass)$^2$ has no sizable and negative
radiative corrections.  Thus, the constraints $m_{\tilde \tau}^2>0$
and $m_{\tilde \tau_1} \ge 72$GeV are no longer serious.  In addition
most cases lead to the neutralino LSP.  Furthermore, the SUSY
contributions to $BR(b \rightarrow s \gamma)$ is roughly proportional
to $\tan \beta$.  Hence, the constraint due to $BR(b \rightarrow s
\gamma)$ is also relaxed for small $\tan \beta$.

\section{Conclusions}

We have studied the large $\tan \beta$ scenario of the SUSY model in
which the gaugino masses are not universal at the GUT scale. We find
that the gluino mass at the electroweak scale is restricted to
multi-TeV values due to experimental limits on the $b
\rightarrow s \gamma$ decay for $\mu < 0$.
In the model {\bf 75} the allowed region is narrow for
$M_3^{\text{GUT}}$.  We find that in two of the models {\bf 1} and
{\bf 24} we have neutralino LSP and stau NLSP, while in the models
{\bf 75} and {\bf 200} the lightest neutralino and the lighter
chargino are almost mass degenerate. This would provide for quite
different kind of the first signature for the MSSM as is usually
assumed within the minimal supergravity scenario. We have also
calculated the SUSY correction to the bottom mass $\delta_b$.  The
model {\bf 75}, as well as the model {\bf 200} with large
$m_0^{\text{GUT}}$, leads to smaller $\delta_b$ than the others.

We have possibilities that gaugino fields acquire a different pattern
of non-universal masses.  For example, there is the case that some
linear combination of $F$-components of {\bf 1}, {\bf 24}, {\bf 75}
and {\bf 200} contributes to gaugino masses.  It is pointed out that
there exists a model-independent contribution to gaugino masses from
the conformal anomaly \cite{GLMR}.  Furthermore, soft scalar masses
and $A$-parameters at the GUT scale can, in general, be non-universal.
We leave these types of extension to future work.

{\bf Note added:}

After completion of this paper, Ref.\cite{newnon} appears, where
several signals of the $SU(5)$ GUTs with non-universal gaugino masses
have been discussed for $\tan \beta = 5$ and 25.

\section*{Acknowledgments}
The authors would like to thank K. {\"{O}}sterberg for useful
correspondence.  This work was partially supported by the Academy of
Finland under Project no. 44129. Y.K. acknowledges support by the
Japanese Grant-in-Aid for Scientific Research ($\sharp$10740111) from
the Ministry of Education, Science and Culture.

%%%%%%%%%%%%%%%%%%%%%%%%%%%%%%%%%%%%%%%%%%%%%%%%%%%%%%%%%%%%%%%%%%%%%%

\begin{table}
\label{tab:gauginos}
\caption{Relative masses of gauginos for different representations of
the $F$-term at the GUT scale and the corresponding relations at the
weak scale. The singlet representation {\bf 1} of the $F$-term
corresponds to the minimal supergravity model.}
\begin{tabular}{|c|ccc|ccc|}
 $F_\Phi$ & $M_1^{\text{GUT}}$ & $M_2^{\text{GUT}}$ &
$M_3^{\text{GUT}}$ & $M_1^{m_Z}$ & $M_2^{m_Z}$ & $M_3^{m_Z}$ \\
\tableline
{\bf 1} & $1$ & $1$ & $1$ & $0.4$ & $0.8$ & $2.9$ \\
{\bf 24} & $-0.5$ & $-1.5$ & $1$ & $-0.2$ & $-1.2$ & $2.9$ \\
{\bf 75} & $-5$ & $3$ & $1$ & $-2.1$ & $2.5$ & $2.9$ \\
{\bf 200} & $10$ & $2$ & $1$ & $4.1$ & $1.6$ & $2.9$ \\
\end{tabular}
\end{table}

\begin{table}
\caption{Mass spectra in the four models 
({\bf 1}, {\bf 24}, {\bf 75}, {\bf 200}) for $\tan \beta=40$.
% $M_3^{\text{GUT}}=800$GeV,
%${\tilde{m}}_0^{\text{GUT}}=400$GeV. 
All the masses are shown in GeV and evaluated at the scale
$m_0^{\text{GUT}}$.}
\label{tab:modellist}
\begin{tabular}{lrrrrrrr}
Model & $\tan \beta$ & $\mu$ & $m_{H^\pm}$ &
$m_{{\tilde{\chi}}^\pm_{1,2}}$ & $m_{{\tilde{\chi}}^0_{1,2,3,4}}$ &
$m_{{\tilde{e}}_{1,2}}$ & $m_{{\tilde{\tau}}_{1,2}}$ \\ 
($M_3^{\text{GUT}},m_0^{\text{GUT}}$)& $\frac{\Gamma (b
\rightarrow s \gamma)}{\Gamma_{SM}}$ & $M_3$ &
$m_{{\tilde{\nu}}_e}/m_{{\tilde{\nu}}_{\tau}}$ & $m_{{\tilde{u}}_{1,2}}$ &
$m_{{\tilde{t}}_{1,2}}$ & $m_{{\tilde{d}}_{1,2}}$ &
$m_{{\tilde{b}}_{1,2}}$ \\
\tableline ${\bf{1}}$ & $40$ & $-982$ &
 $472$ &
 $660/993$ &
 $342/660/985/993$ & $506/690$ &
 $407/676$ \\
(800, 400) & $1.5$ & $1963$ & $685/658$ & $1749/1819$ &
 $1375/1573$ & $1740/1821$ & $1468/1561$ \\
\tableline ${\bf{24}}$ & $40$ & $-791$ &
 $581$ &
 $778/1018$ &
 $170/778/794/1018$ & $430/906$ &
 $220/872$ \\
(800, 400) & $1.4$ & $1963$ & $903/865$ & $1740/1916$ &
 $1394/1681$ & $1738/1917$ & $1517/1674$ \\
\tableline ${\bf{75}}$ & $40$ & $-8$ &
 $1521$ &
 $8/2006$ &
 $7/9/1717/2006$ & $1592/1836$ &
 $1387/1751$ \\
(800, 400) & $1.2$ & $1963$ & $1834/1749$ & $2018/2387$ &
 $1287/2072$ & $1812/2388$ & $1668/2061$ \\
\tableline ${\bf{200}}$ & $40$ & $-784$ &
 $1218$ &
 $780/1342$ &
 $778/785/1342/3433$ & $1923/3107$ &
 $1721/2861$ \\
(800, 400) & $1.2$ & $1963$ & $1921/1720$ & $2108/2690$ &
 $1480/2053$ & $2018/2109$ & $1480/1642$ \\
\tableline ${\bf{1}}$ & $40$ & $-464$ &
 $1017$ &
 $82/477$ &
 $44/82/471/473$ & $1501/1502$ &
 $1260/1388$ \\
(100, 1500) & $1.3$ & $222$ & $1500/1385$ & $1511/1512$ &
 $830/1065$ & $1511/1514$ & $1053/1247$ \\
\tableline ${\bf{24}}$ & $40$ & $-457$ &
 $1013$ &
 $122/472$ &
 $21/122/464/469$ & $1500/1503$ &
 $1259/1390$ \\
(100, 1500) & $1.3$ & $222$ & $1501/1387$ & $1511/1513$ &
 $830/1066$ & $1511/1515$ & $1055/1245$ \\
\tableline ${\bf{75}}$ & $40$ & $-444$ &
 $1027$ &
 $243/463$ &
 $219/243/453/460$ & $1512/1516$ &
 $1271/1402$ \\
(100, 1500) & $1.3$ & $222$ & $1514/1399$ & $1516/1523$ &
 $828/1077$ & $1513/1525$ & $1066/1247$ \\
\tableline ${\bf{200}}$ & $40$ & $-456$ &
 $1032$ &
 $164/471$ &
 $164/423/462/488$ & $1518/1548$ &
 $1308/1403$ \\
(100, 1500) & $1.3$ & $222$ & $1516/1399$ & $1517/1532$ &
 $850/1066$ & $1517/1519$ & $1055/1252$ \\
\end{tabular}
\end{table}

\begin{figure}
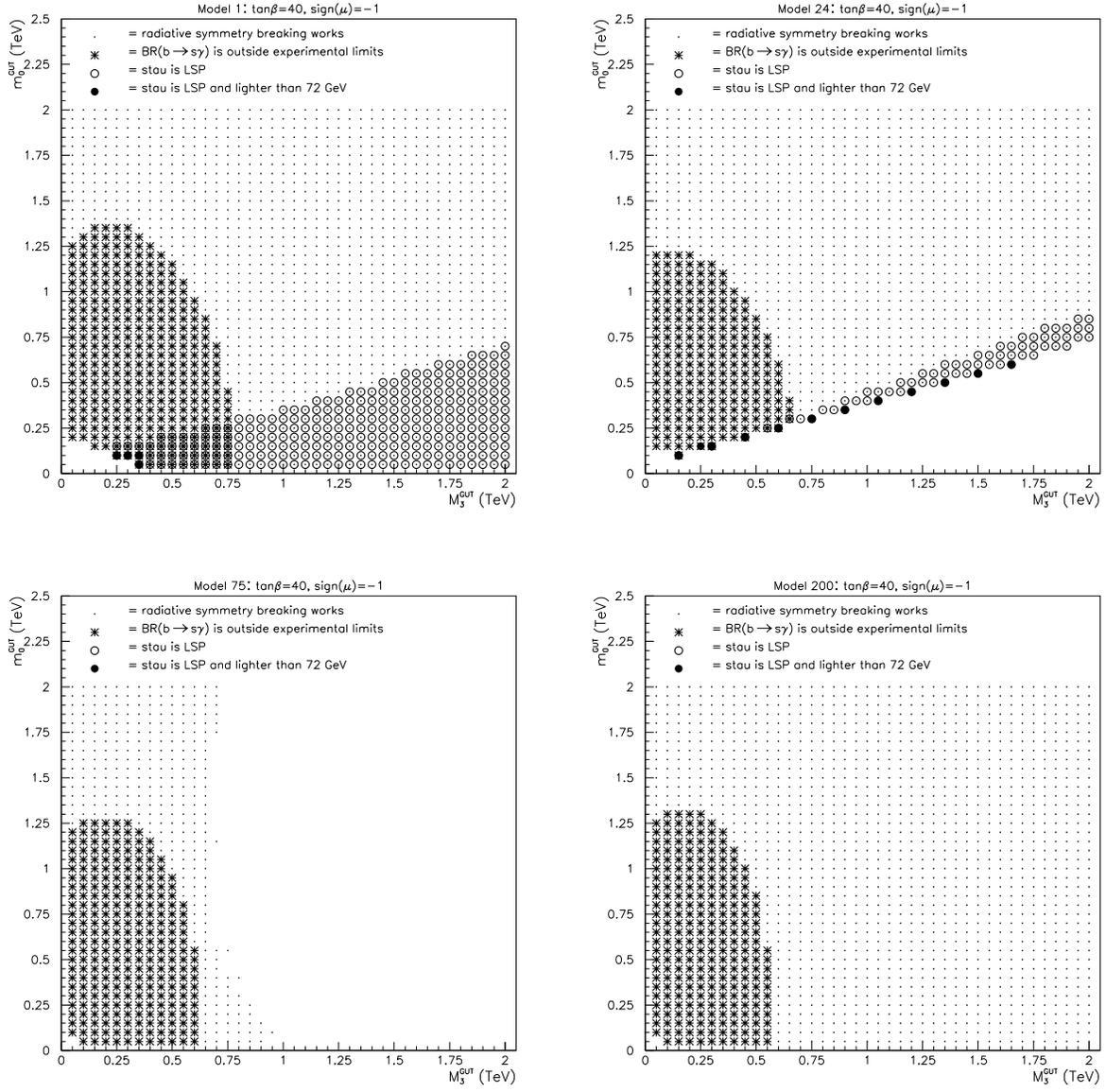

\label{fig:1}
\caption{Scan over the gluino mass term $M_3^{GUT}$ and the universal
scalar mass term $m_0^{GUT}$  for all four models ({\bf 1},
{\bf 24}, {\bf 75}, {\bf 200}; $\tan \beta=40$).}
$\begin{array}{cc}
\mbox{\epsfxsize=8cm \epsfysize=8cm \epsffile{itku-1n.epsi}} & 
\mbox{\epsfxsize=8cm \epsfysize=8cm \epsffile{itku-24n.epsi}} \\
\mbox{\epsfxsize=8cm \epsfysize=8cm \epsffile{itku-75n.epsi}} &
\mbox{\epsfxsize=8cm \epsfysize=8cm \epsffile{itku-200n.epsi}}
\end{array}$
\end{figure}

\newpage
\begin{figure}
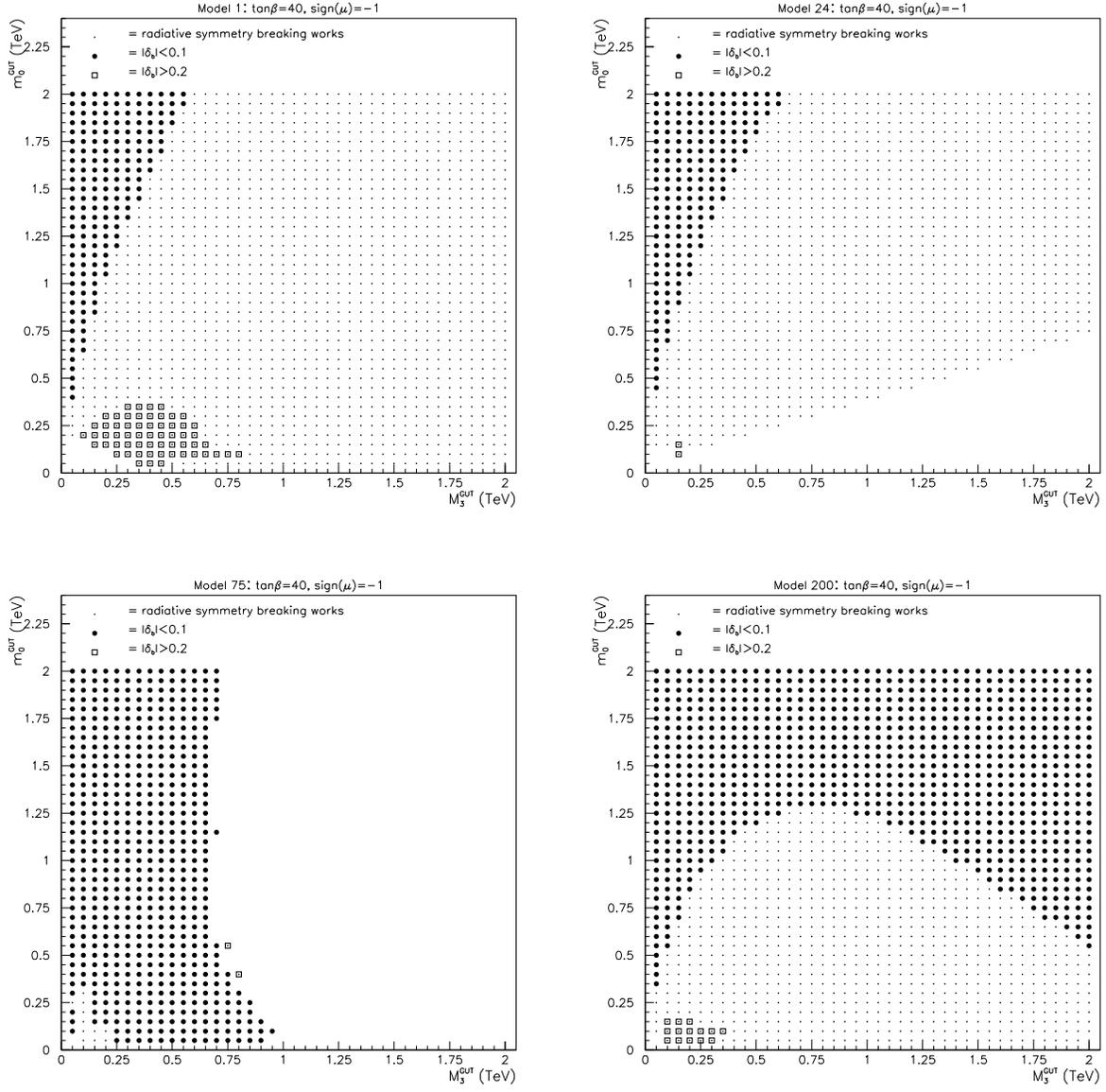

\label{fig:2}
\caption{The bottom mass correction $|\delta_b|$ 
for all four models ({\bf 1},
{\bf 24}, {\bf 75}, {\bf 200}; $\tan \beta=40$)}
$\begin{array}{cc}
\mbox{\epsfxsize=8cm \epsfysize=8cm
\epsffile{lukeekohankukaantata-1nn.epsi}} &
\mbox{\epsfxsize=8cm \epsfysize=8cm
\epsffile{lukeekohankukaantata-24nn.epsi}} \\
\mbox{\epsfxsize=8cm \epsfysize=8cm
\epsffile{lukeekohankukaantata-75nn.epsi}} &
\mbox{\epsfxsize=8cm \epsfysize=8cm \epsffile{lukeekohankukaantata-200nn.epsi}}
\end{array}$
\end{figure}

\end{document}